\documentclass{article}
\usepackage{amssymb}

\input{tcilatex}
\begin{document}

\title{ \textbf{\ Massive Neutrinos, Lolrentz Invariance Dominated Standard
Model and the Phenomenological Approach to Neutrino Oscillations }\\
(Phys. Scr.80 (2009) 025101)}
\author{Josip \v{S}oln \\
%EndAName
Army Research Laboratory (ret.),\\
JZS Phys-Tech\\
Vienna, Virginia 22182, USA\\
soln.phystech@cox.net}
\maketitle

\textbf{.}

\bigskip

\textbf{\ Abstract}

For the electroweak interactions, the massive neutrino perturbative
kinematical procedure is developed in the massive neutrino Fock space; The
perturbation expansion parameter is the ratio of neutrino mass to its
energy. This procedure, within the Pontecorvo-Maki-Nakagawa-Sakata (PMNS)
modified electroweak Lagrangian, calculates the cross-sections with the new
neutrino energy projection operators in the massive neutrino Fock space,
resulting in the dominant Lorentz invariant Standard Model mass-less flavor
neutrino cross-sections. As a consequence of the kinematical relations
between the massive and massless neutrinos,some of the neutrino oscillation
cross-sections are Lorentz invariance violating. But \ all these oscillating
cross-sections, some of which violate the flavor conservation, being
proportional to the squares of neutrino masses are practically unobservable
in the laboratory. However, these neutrino oscillating cross-sections are
consistent with the original Pontecorvo neutrino oscillating transition
probability expression at short time (baseline), as presented by Dvornikov.
From these comparisons, by mimicking the time dependence of the original
Pontecorvo neutrino oscillating transition probability, one can formulate
the dimensionless neutrino intensity-probability $I$, by phenomenologicaly
extrapolating of the time $t$, or, equivalently the baseline distance $L$
away from the collision point for the oscillating differential
cross-section. For the incoming neutrino of $10MeV$ in energy and neutrino
masses from Fritzsch analysis with the neutrino mixing matrix of Harrison,
Perkins and Scott, the baseline distances at the first two maxima of the
neutrino intensity are $L\simeq 281km$ and $L\simeq 9279km$ . The intensity $%
I$ at the first maximum conserves the flavor, while at the second maximum,
the intensities violate the flavor, respectively, in the final and initial
state. At the end some details are given as to how these neutrino
oscillations away from the collision point one should be able to verify
experimentally. In material presented here one reinforces the notion that
the mass-less flavor neutrino can be considered as the superposition of
three massive neutrinos.

\bigskip

\textbf{\ Introduction}

The flavor changing neutrino oscillations experiments, such as, The
Super-Kamiokande [1], SNO [2], KAMLAND [3] as well as Homestake
Collaboration [4], clearly require massive neutrinos as have been exhibited,
for example, by Bilenky, Giunti and Grimus [5] , Giunti and Laveder [6] and
Kayser [7]. In discussing the neutrino oscillations, one assumes that the
left-handed flavor mass-less neutrino fields $\nu _{\alpha L}$, with $\alpha
=e,\mu ,\tau $, are unitary linear combinations of of the massive neutrino
fields $\nu _{iL}$and analogously for the states (see [5-10])and references
therein),

\begin{equation}
\nu _{\alpha L}=U_{\alpha i}\nu _{iL},\mid \nu _{\alpha }\rangle =U_{i\alpha
}^{\dagger }\mid \nu _{i}\rangle \ \ (i=1,2,3;\alpha =e,\mu ,\tau )\ \ \ 
\text{\ \ \ }\ \ \ \   \tag{1,2}
\end{equation}

\ \ \ \ \ \ \ \ \ \ \ \ \ \ \ \ \ \ \ \ \ \ \ \ \ \ \ \ \ \ \ \ \ \ \ \ \ \
\ \ \ \ \ \ \ \ \ \ \ \ \ \ \ \ \ \ \ \ \ \ \ \ \ \ \ \ \ \ \ \ \ \ \ \ \ \
\ \ \ \ \ \ \ \ \ \ \ \ \ \ \ \ \ \ \ \ \ \ \ \ \ \ \ \ \ \ \ \ \ \ \ \ \ \
\ \ \ \ \ \ \ \ \ \ \ \ \ \ \ \ \ \ \ \ \ \ \ \ \ \ \ \ \ \ \ \ \ \ \ \ \ \
\ \ \ \ \ \ \ \ \ \ \ \ \ \ \ \ \ \ \ \ \ \ \ \ \ \ \ \ \ \ \ \ \ \ \ \ \ \
\ \ \ \ \ \ \ \ \ \ \ \ \ \ \ \ \ \ \ \ \ \ \ \ \ \ \ \ \ \ \ \ \ \ \ \ \ \
\ \ \ \ \ \ \ \ \ \ \ \ \ \ \ \ \ \ \ \ \ \ \ \ \ \ \ \ \ \ \ \ \ \ \ \ \ \
\ \ \ \ \ \ \ \ \ \ \ \ \ \ \ \ \ \ \ \ \ \ \ \ \ \ \ \ \ \ \ \ \ \ \ \ \ \
\ \ \ \ \ \ \ \ \ \ \ \ \ \ \ \ \ \ \ \ \ \ \ \ \ \ \ \ \ \ \ \ \ \ \ \ \ \
\ \ \ \ \ \ \ \ \ \ \ \ \ \ \ \ \ \ \ \ \ \ \ \ \ \ \ \ \ \ \ \ \ \ \ \ \ \
\ \ \ \ \ \ \ \ \ \ \ \ \ \ \ \ \ \ \ \ \ \ \ \ \ \ \ \ \ \ \ \ \ \ \ \ \ \
\ \ \ \ \ \ \ \ Here $U$ is the unitary Pontecorvo-Maki-Nakagawa-Sakata
(PMNS) matrix and $\nu _{iL}$\ \ is a left-handed neutrino field associated
with mass $m_{i}$ (see, for example [5-11]). In the study of flavor neutrino
oscillations, the flavor state from (2) is used, via the Schr\"{o}dinger
equation, to calculate the oscillation probability $P_{\alpha \beta }(L)$ \
for the flavor neutrino oscillation transition $\nu _{\alpha
}\longrightarrow $\ $\nu _{\beta }$\ at a very large distance $L(\backsim
thousands$ $of$ $km)$\ (see, for example [5-11] ).

Not long after the suggestion by Pontecorvo [12] that massive neutrinos
oscillate, which appears is the easiest to treat with the Schroedinger
equation (consult [5-11]), people started looking as to how the neutrino
masses influence the field theoretic treatment of the standard model (SM).
One of the earliest paper that addressed that problem was by R. E. Schrock
[13]. He studied mostly the decays where the neutrinos and antineutrinos
appear in final states, such as the $\pi ,K$ and the nuclear $\beta $
decays. Such decays are friendly to the application of the PMNS unitary
transformations to the established SM. Here, the kinematics is then of the
massive neutrinos augmented with the PMNS unitary matrix dependence. He
proposed the tests that showed the promise for determination of neutrino
masses and \ lepton mixing. Specifically, the proposed tests of the $\pi $
and $K$ decays could detect the neutrino masses in the 1 to 400 MeV range
while the nuclear $\beta $ decays in the 0.1 keV to 5 MeV range. Today it is
known that all these ranges are too high as it appears that every neutrino
mass is bellow 1 eV.

More recently, Li and Liu [14] have studied the connection of massive
neutrinos to the SM by studying the inequivalent vaccua model (see [15] and
references therein); here, the transformation between the Fock space of
neutrino mas states and the unitary inequivalent flavor states is \ a
Bogoliubov transformation [14], [15]. This transformation, for instance,
yields the corrections to the Pontecorvo neutrino oscillation probability
that are of $O(m^{2})$ $(m$ denoting generically any of the three neutrino
masses). However, the problem is that it also yields that the branching
ration of $W^{+}\rightarrow e^{+}+\nu _{\mu }$ to $W^{+}\rightarrow
e^{+}+\nu _{e}$ is of $O(m^{2})$ which contradicts the Hamiltonian that one
started from. In other words, in the inequivalent vaccua model, there is a
flavor changing current such as $W^{+}\rightarrow e^{+}+\nu _{\mu }$ and the
branching ratio is different from that the SM with zero neutrino masses. In
Appendix, Li and Liu show that the neutrino oscillation effects are large
enough to neglect the inequivalent vaccua model effects; that is, the sum of
all three decay widths of $W^{+}\rightarrow e^{+}+\nu _{e,\mu ,\tau }$
equals the width of $W^{+}\rightarrow e^{+}+\nu _{e}$ in the SM.

It is well known that all by itself, the standard model with mass-less
flavor neutrinos has been remarkably successful in describing the laboratory
experimental data, such as the neutrino scattering, at low and medium
energies ( see, for instance, M. Fukugita and T. Yanagida [8]; and C. Giunti
and C. W. Kim [9]). So one can then ask whether the SM cross-sections can be
derived when starting, instead with the mass-less flavor neutrino fields \ \ 
$\nu _{\alpha L}$, with the massive neutrino fields \ $\nu _{iL}$. In this
article the answer to this question is in affirmative.

To proceed in this direction, as suggested by (1), the application of the
PMNS substitution rule (3) transforms the SM Lagrangian density with the
mass-less neutrino fields into the one with the massive neutrino fields
(4):\ \ $\ $

$\ \ \ \ \ \ \ \ \ \ \ \ \ \ \ \ \ \ \ \ \ \ \ \ \ \ \ \ \ \ \ \ \ \ \ \ \ \
\ \ \ \ \ \ \ \ \ \ \ \ \ \ \ \ \ \ \ \ \ \ \ \ \ \ \ \ \ \ \ \ \ \ \ \ \ \
\ \ \ \ \ \ \ \ \ \ \ \ \ \ \ \ \ \ \ \ \ \ \ \ \ \ \ \ \ \ $%
\begin{equation}
The\text{ }PMNS\text{ }substitution:\nu _{\alpha L}\rightarrow U_{\alpha
i}\nu _{iL}  \tag{3}
\end{equation}%
\begin{eqnarray}
\alpha ,\beta ,...,\epsilon &=&e,\mu ,\tau ;i,j,a,...,b=1,2,3:  \nonumber \\
l_{\alpha L} &=&\left( 
\begin{array}{c}
U_{\alpha i}\nu _{iL} \\ 
\alpha _{L}%
\end{array}%
\right) ,\epsilon _{L,R}=P_{L,R}\epsilon ,P_{L,R}=\frac{1}{2}\left( 1\mp
\gamma ^{5}\right)  \nonumber \\
L_{W,int}^{Lepton} &=&\frac{g}{\sqrt{2}}\dsum\limits_{\epsilon =e,\mu ,\tau
;i=1,2,3}[\overline{\nu }_{iL}(x)U_{i\epsilon }^{\dagger }\gamma ^{\mu
}\epsilon _{L}(x)W_{\mu }(x,+)  \nonumber \\
&&+\overline{\epsilon }_{L}(x)\gamma ^{\mu }U_{\epsilon j}\nu _{jL\left(
x\right) }W_{\mu }^{\dagger }(x,+)],  \TCItag{4} \\
W^{\mu }(x,\pm ) &=&\frac{1}{\sqrt{2}}\left[ W^{\mu }(x,1)\mp iW^{\mu }(x,2)%
\right]  \nonumber \\
L_{Z,int}^{Lepton} &=&\frac{g}{c_{W}}Z_{\mu }(x)\dsum\limits_{\epsilon
=e,\mu ,\tau }\left[ \overline{l}_{\epsilon L}(x)\frac{\tau _{3}}{2}\gamma
^{\mu }l_{\epsilon L}(x)-s_{W}^{2}(-)\overline{\epsilon }(x)\gamma ^{\mu
}\epsilon (x)\right]  \nonumber \\
&=&\frac{g}{4c_{W}}Z_{\mu }(x)\dsum\limits_{\epsilon =e,\mu ,\tau
,a,b=1,2,3}[\overline{\nu _{a}}(x)U_{a\epsilon }^{\dagger }\gamma ^{\mu
}\left( 1-\gamma ^{5}\right) U_{\epsilon b}\nu _{b}(x)  \nonumber \\
&&+\overline{\epsilon }(x)\gamma ^{\mu }\left[ \left( 4s_{W}^{2}-1\right)
+\gamma ^{5}\right] \epsilon (x)],  \nonumber \\
s_{W} &=&\sin \theta _{W},c_{W}=\cos \theta _{W\text{ \ \ \ \ \ \ \ \ \ \ \
\ \ \ \ \ \ \ \ \ \ \ \ \ \ \ \ \ \ \ \ \ \ \ \ \ \ \ \ \ \ \ \ \ \ \ \ \ \
\ \ \ \ \ \ \ \ \ \ \ \ \ \ \ \ \ \ }}  \nonumber
\end{eqnarray}

\ \ \ 

Since the Lagrangian densities(4) contains the massive neutrino fields, all
the calculations are now done formally in the massive neutrino Fock space.
The mass-less neutrinos will be the mass state neutrinos in the limit of
negligible masses as a result of the perturbative neutrino kinematical
procedure.

\bigskip

\textbf{Perturbative kinematical procedure for calculating the neutrino
differential cross-sections}

\bigskip

A free neutrino spinor field with the mass $m_{i}$,$i=1,2,3$ , is written
generally with the creation and annihilation operators as

\begin{eqnarray}
\nu _{i}(x) &=&\frac{1}{\left( 2\pi \right) ^{\frac{3}{2}}}\int \frac{d^{3}q%
}{q^{0}}\sum_{s}e^{iqx}u(q,s)a(q,s)+e^{-iqx}v(q,s)b^{\dagger }(q,s) 
\nonumber \\
q^{0} &=&\left( \overrightarrow{q}^{2}+m_{i}^{2}\right) ^{\frac{1}{2}}\text{
\ \ \ \ \ \ \ \ \ \ \ \ \ \ \ \ \ \ \ \ \ \ \ \ \ \ \ \ \ \ \ \ \ \ \ \ \ \
\ \ \ \ \ \ \ \ \ \ \ \ \ \ \ \ \ }  \TCItag{5}
\end{eqnarray}

The perturbative kinematics is based on the fact that the neutrino mass \ $%
m_{i}(m_{i}\lessdot 1eV)$is generally much smaller than its absolute
momentum value$\left\vert \overrightarrow{q}\right\vert $. Therefore it is
convenient to start with the \textquotedblleft mass-less\textquotedblright\
four-component neutrino momentum \ $q_{(\gamma )}^{\mu }$ with fixed flavor
parameter $\gamma $

\begin{equation}
q_{\left( \gamma \right) }^{\mu }=\left( \overrightarrow{q}_{\left( \gamma
\right) },q_{\left( \gamma \right) }^{0}\right) ,q_{\left( \gamma \right)
}^{2}=0,\gamma =e,\mu ,\tau \text{ \ \ \ \ \ }  \tag{6}
\end{equation}%
Next, one assumes that under this flavor parameter $\gamma $ are grouped
together three massive neutrinos, say, $\nu _{i}$ with masses $m_{i}$ $%
;i=1,2,3$ then the difference among their energies $\left\vert \Delta
q_{(i_{1},i_{2})}^{0}\right\vert =\left\vert
(q_{(i_{2})}^{0}-q_{(i_{1})}^{0})\right\vert \cong \left\vert \Delta
m_{i_{2,}i_{1}}^{2}\diagup q_{(\gamma )}^{0}\right\vert =\left\vert
(m_{i_{2}}^{2}-m_{i_{1}}^{2})\diagup q_{(\gamma )}^{0}\right\vert $ is much
smaller than the quantum-mechanical uncertainty of the energy [16]. As a
consequence, in this case with fixed $\gamma $ it is impossible to
distinguish the emission of neutrinos with different masses in the neutrino
processes [16]. Hence, the three massive neutrinos, satisfying these quantum
mechanical conditions, can be viewed as superposing themselves to form the
flavor neutrino $\nu _{\gamma }$ [11,16] as depicted by relations (1) and
(2). With this in mind, with $\ q_{(i,\gamma )}^{\mu }$ $=\left( 
\overrightarrow{q}_{(\gamma )},\left( \overrightarrow{q}_{(\gamma
)}^{2}+m_{i}^{2}\right) ^{\frac{1}{2}}\right) $as the four-momentum of the
massive neutrino with mass $m_{i}$ the perturbative kinematics can be
presented as

\begin{eqnarray}
q_{(i,\gamma )}^{\mu } &\simeq &q_{(\gamma )}^{\mu }-g^{\mu 0}\frac{m_{i}^{2}%
}{2q_{(\gamma )}^{0}}+O\left( m_{i}^{4}\right) ,i=1,2,3;\gamma
(fixed)=(e,\mu ,\tau );  \nonumber \\
\overrightarrow{q}_{\left( i,\gamma \right) } &=&\overrightarrow{q}_{\left(
\gamma \right) },q_{(i,\gamma )}^{0}\simeq q_{(\gamma )}^{0}+\frac{m_{i}^{2}%
}{2q_{(\gamma )}^{0}}+O\left( m_{i}^{4}\right) ,  \TCItag{7} \\
q_{(i,\gamma )}^{2} &\simeq &-m_{i}^{2}\text{ }+O\left( m_{i}^{4}\right) 
\text{\ \ \ \ \ }  \nonumber
\end{eqnarray}%
In (7) the terms with $O(m_{i}^{4})$ will be neglected and the fixed
parameter $\gamma $, as already established is the neutrino flavor. Thus
with this perturbative kinematics $q_{\left( i,\gamma \right) }^{\mu }$%
ceases to be a true Lorentz four-momentum, while $q_{\left( \gamma \right)
}^{\mu }$ remains to be so. However, since $q_{\left( \gamma \right) }^{\mu
} $, as shown in (7), is the main part of $q_{\left( i,\gamma \right) }^{\mu
}$, the main portions of cross-sections are expected to be Lorentz invariant 
$(LI)$ while the Lorentz invariance violation $(LIV)$ will be generated by $%
\ g^{\mu 0}m_{i}^{2}/2q_{\left( \gamma \right) }^{0}$ from (7). The $LIV$
terms are expected to be very small due the smallness of neutrino masses.
Taking these relations into account, within the massive neutrino Fock space
the differential cross-sections with flavor neutrinos are calculated. The
question, of course is: is the result consistent with the SM?

To continue, in analogy to $q_{(i,\gamma )}^{\mu }$, one now introduces $%
\widehat{s}_{(i,\gamma )}$ and $s_{(i,\gamma )}$ to denote respectively, the
helicity operators and eigenvalues for \ $i=1,2,3$ massive neutrinos
comprising the mass-less flavor neutrino $\nu _{\gamma }$; The helicity
operator and eigenvalue of the mass-less flavor neutrino $\ \nu _{\gamma }$
are denoted, respectively, as $\widehat{s}_{(\gamma )}$ and $s_{(\gamma )%
\text{ }}$. And, the effects of the massive to mass-less-neutrino
kinematical relation (7) on these helicity eigenvalues are simply, what one
can call, the ordinary massive to mass-less neutrino helicity relation.

\begin{eqnarray}
\widehat{s}_{(i,\gamma )} &=&\overrightarrow{q}_{\left( i,\gamma \right)
}\cdot \overrightarrow{\sigma }\diagup \downharpoonright \overrightarrow{q}%
_{\left( i,\gamma \right) }\downharpoonright =\overrightarrow{q}_{\left(
\gamma \right) }\cdot \overrightarrow{\sigma }\diagup \downharpoonright 
\overrightarrow{q}_{\left( \gamma \right) }\downharpoonright =\widehat{s}%
_{(\gamma )},  \nonumber \\
\widehat{s}_{(i,\gamma )} &=&\widehat{s}_{(\gamma )}=\widehat{s}_{(k,\gamma
)}\Longrightarrow s_{(i,\gamma )}=s_{(\gamma )}=s_{(k,\gamma )},etc; 
\TCItag{8} \\
i\text{ }or\text{ }k,... &=&1,2,3;\gamma (fixed)=(e\text{ }or\text{ }\mu 
\text{ }or\text{ }\tau )  \nonumber
\end{eqnarray}

As a consequence of (7) and (8), with spinor indices suppressed, the
contractions of massive neutrino free-field operators with the massive
neutrino and antineutrino states are, respectively

\begin{eqnarray}
\left\langle 0\right\vert \text{\ }\nu \left( x,l\right) \left\vert
q_{\left( i,\gamma \right) },s_{\left( i,\gamma \right) }\right\rangle \text{%
\ } &=&\frac{1}{\left( 2\pi \right) ^{\frac{3}{2}}}\text{\ }e^{i\left(
q_{\left( i,\gamma \right) }\cdot x\right) }\delta _{li}u\left( q_{\left(
i,\gamma \right) },s_{_{\left( i,\gamma \right) }}\right) ,\text{\ \ \ \ \ }
\nonumber \\
\left\langle \overline{q}_{\left( j,\delta \right) },s_{\left( j.\delta
\right) }\right\vert \text{\ }\nu \left( x,k\right) \text{\ }\left\vert
0\right\rangle &=&\frac{1}{\left( 2\pi \right) ^{\frac{3}{2}}}\text{\ }%
e^{-i\left( q_{\left( k,\delta \right) }\cdot x\right) }\delta _{kj}v\left(
q_{\left( j,\delta \right) },s_{_{\left( j,\delta \right) }}\right) \text{\
\ \ \ \ \ \ \ \ \ \ \ \ \ \ \ \ \ \ \ \ \ \ \ }  \TCItag{9}
\end{eqnarray}%
where $s_{(j,\delta )}$ and $\delta $ have the same kind interrelationship \
as $s_{(i,\gamma )}$ and $\gamma $ in (8), etc.

Since , as shown in (7) to (9), the superposed three massive neutrinos
contain the single flavor designation, either in the initial or final state,
say, $\gamma $ and $\delta $, the process can be denoted as $\ \nu (\gamma
)+\alpha (P_{1})\rightarrow \nu (\delta )+\beta (P_{2})$ .From the
Lagrangian densities (4) the amplitude and its Hermitian conjugate for the
process containing massive neutrinos, are build around these respective
flavor designations, $\gamma $ and $\delta $. so that the generic amplitudes
are given, respectively, as

\begin{eqnarray}
S_{amp} &\backsim &\sum_{i,j,...}\delta _{4}(q_{(i,\gamma
)}+P_{(1)}-q_{(j,\delta )}-P_{(2)})iM_{i,j,...},  \nonumber \\
S_{amp}^{\dagger } &\backsim &\sum_{k,l,...}\delta _{4}(q_{(k,\gamma
)}+P_{(1)}-q_{(l,\delta )}-P_{(2)})(-i)M_{k,l,...}^{\ast }  \TCItag{10}
\end{eqnarray}%
Here, the momenta indicate the actual massive neutrino-lepton scattering and
different Latin indices indicate possibilities of summation with the $U$
matrices which, however, here is not necessary to be explicit. To derive the
cross-section, with the help of (7), (8) and (9), one needs

\begin{eqnarray}
S_{amp}^{\ast }S_{amp} &=&\sum_{i.j,...;k,l,...}\left\{ \delta
_{4}^{2}(q_{(\gamma )}+P_{(1)}-q_{(\delta )}-P_{(2)})\right.  \nonumber \\
&&+\delta _{3}^{2}(\overrightarrow{q}_{(\gamma )}+\overrightarrow{P}_{(1)}-%
\overrightarrow{q}_{(\delta )}-\overrightarrow{P}_{(2)})\delta (q_{(\gamma
)}^{0}+P_{(1)}^{0}-q_{(\delta )}^{0}-P_{(2)}^{0})  \nonumber \\
&&\times \delta ^{\prime }(q_{(\gamma )}^{0}+P_{(1)}^{0}-q_{(\delta
)}^{0}-P_{(2)}^{0})\frac{1}{2}\left[ \frac{m_{i}^{2}+m_{k}^{2}}{q_{(\gamma
)}^{0}}-\frac{m_{j}^{2}+m_{l}^{2}}{q_{(\delta )}^{0}}\right]  \TCItag{11} \\
&&\left. +O(m^{4})\right\} (M_{i,j,...})(M_{k,l,...}^{\ast })  \nonumber \\
&=&\delta _{4}^{2}(q_{(\gamma )}+P_{(1)}-q_{(\delta
)}-P_{(2)})\sum_{i,j,...;k,l,...}(M_{i,j,...})(M_{k,l,...}^{\ast })+O\left(
m^{4}\right) \text{\ }  \nonumber
\end{eqnarray}%
The final result in (11) is the consequence of general delta function
property $\delta (x)\delta ^{\prime }(x)=0$ \ .The terms with $O\left(
m^{4}\right) $ , denoting the fourth power of products of variety of $%
m_{i},m_{k},etc.$, are neglected. It follows that while the Fock space
contains the massive neutrino states, the cross-section will utilize the
kinematics of massless flavor neutrinos.

Next, one needs the spinor expressions, appearing in (9), to reflect
respectively, the kinematical and helicity relations in order to facilitate
the cross-section calculations.%
\begin{eqnarray}
u(q_{(i,\alpha )},s_{(i,\alpha )}) &=&\frac{m_{i}-\underline{q}_{(i,\alpha )}%
}{\sqrt{2\left( m_{i}+q_{(i,\alpha )}^{0}\right) }}u(m_{i},\overrightarrow{0}%
,s_{(i,\alpha )}),  \nonumber \\
u(m_{i},\overrightarrow{0},s_{(i,\alpha )} &=&\pm 1)=\left( 
\begin{array}{c}
1 \\ 
0 \\ 
0 \\ 
0%
\end{array}%
\right) ,\left( 
\begin{array}{c}
0 \\ 
1 \\ 
0 \\ 
0%
\end{array}%
\right) ,  \TCItag{12} \\
\underline{q}_{(i,\alpha )} &=&\gamma _{\mu }q_{(i,\alpha )};  \nonumber \\
\overline{u}(q_{(i,\alpha )},s_{(i,\alpha )}) &=&\overline{u}(m_{i},%
\overrightarrow{0},s_{(i,\alpha )})\frac{m_{i}-\underline{q}_{(i,\alpha )}}{%
\sqrt{2\left( m_{i}+q_{(i,\alpha )}^{0}\right) }}  \nonumber
\end{eqnarray}

For a process $\ $with $\gamma $ and $\delta $ flavor designations, $\ \nu
(\gamma )+\alpha (P_{1})\longrightarrow \nu (\delta )+\beta (P_{2})$, in
cross-section evaluations, one will deal with the neutrino energy projection
operator over the positive energy states. Furthermore, rather than averaging
over, one simply sums over the massive neutrino helicity degrees of freedom.
Consistent with the ordinary neutrino helicity relation (8), the sum is
carried over only the equal helicity eigenvalues:

\begin{eqnarray}
s_{(i,\alpha )} &=&s_{(k,\alpha )}=s_{(\alpha )}:\sum_{s_{(i,\alpha
)},s_{(k,\alpha )}}u(q_{(i,\alpha )},s_{(i,\alpha )})\otimes \overline{u}%
(q_{(k,\alpha )},s_{(k,\alpha )})  \nonumber \\
&=&\sum_{s_{(\alpha )}}u(q_{(i,\alpha )},s_{(\alpha )})\otimes \overline{u}%
(q_{(k,\alpha )},s_{(\alpha )})\equiv \frac{1}{2}\left[ q_{(i,\alpha
)},q_{(k,\alpha )};+,c\right] ,  \nonumber \\
\left[ q_{(i,\alpha )},q_{(k,\alpha )};+,c\right] &=&\frac{\left( m_{i}-%
\underline{q}_{(i,\alpha )}\right) \left( 1+\gamma ^{0}\right) \left( m_{k}-%
\underline{q}_{(k,\alpha )}\right) }{2\left[ \left( m_{i}+q_{(i,\alpha
)}^{0}\right) \left( m_{k}+q_{(k,\alpha )}^{0}\right) \right] ^{\frac{1}{2}}}%
,  \TCItag{13} \\
i\text{ }and\text{ }k &=&1,2,3;\text{ }\gamma =e\text{ }or\text{ }\mu \text{ 
}or\text{ }\tau  \nonumber
\end{eqnarray}%
where the $+$ sign refers to the positive energy states and $c$ refers to
the fact that the equal helicity eigenvalues in the sum yield the coherent
result. (The incoherent projection operators \ $\left[ q_{(i,\alpha
)},q_{(k,\alpha )};+,i\right] $ with unequal helicity eigenvalues \ \ $%
s_{(i,\alpha )}\neq s_{(k,\alpha )}$\ \ \ are not dealt here.) The relation
(13) defines the spinorial massive neutrino to mass-less neutrino helicity
relation and it is consistent with the ordinary helicity relation (8).
Carrying out the indicated operations in relation (13) as a power series
over the neutrino masses /energy, one obtains for the neutrino energy
projection operator over the positive energy states the following

\begin{eqnarray}
\left[ q_{(i,\alpha )},q_{(k,\alpha )};+,c\right] &=&\sum_{n=0}^{2}\left[
q_{(i,\alpha )},q_{(k,\alpha )};+,c\right] _{n},  \nonumber \\
\left[ q_{(i,\alpha )},q_{(k,\alpha )};+,c\right] _{0} &=&-\underline{q}%
_{\left( \alpha \right) },  \TCItag{14} \\
\left[ q_{(i,\alpha )},q_{(k,\alpha )};+,c\right] _{1} &=&m_{k}+\frac{\left(
m_{k}-m_{i}\right) \gamma ^{0}\underline{q}_{\left( \alpha \right) }}{%
2q_{\left( \alpha \right) }^{0}},  \nonumber \\
\left[ q_{(i,\alpha )},q_{(k,\alpha )};+,c\right] _{2} &=&-\frac{\left(
m_{k}-m_{i}\right) ^{2}\underline{q}_{\left( \alpha \right) }}{8q_{\left(
\alpha \right) }^{02}}+\frac{m_{i}m_{k}\gamma ^{0}}{2q_{\left( \alpha
\right) }^{0}}  \nonumber
\end{eqnarray}

The coherent energy operator \ $\left[ q_{(i,\alpha )},q_{(k,\alpha )};+,c%
\right] $ generates the electroweak interactions that are the same as the SM
interactions plus the LIV neutrino oscillation processes that are negligible
since their cross-sections are proportional to the squares of neutrino
masses and, as such, are essentially zero. Relation (14) is in essence the
procedure for calculating the cross-sections for the processes requiring
only the neutrino energy projection operators over the positive energy
states.

\bigskip

\textbf{Applications to the differential cross-section calculations}

\bigskip

As established earlier and consistent with (11), the quasi-elastic
electroweak process with massive neutrinos present, to $O(m^{2})$, can can
be denoted with the kinematics that uses just the mass-less flavor neutrinos.

\begin{equation}
\nu \left( q_{\left( \gamma \right) }\right) +\alpha \left( P_{\left(
1\right) }\right) \longrightarrow \nu \left( q_{\left( \delta \right)
}\right) +\beta \left( P_{\left( 2\right) }\right) ;y=\frac{q_{\left( \gamma
\right) }^{0}-q_{\left( \delta \right) }^{0}}{q_{\left( \gamma \right) }^{0}}%
=\frac{P_{\left( 2\right) }^{0}-P_{\left( 1\right) }^{0}}{q_{\left( \gamma
\right) }^{0}}  \tag{15}
\end{equation}%
where $y$ is the normalized energy transfer. Now, although working in the
massive neutrino Fock space, relation (11) says that the kinematics for the
cross-sections for the quasi-elastic scattering of the massless flavor
neutrinos is determined with flavor neutrino momenta according to $\delta
_{4}\left( q_{\left( \gamma \right) }+P_{\left( 1\right) }-q_{\left( \delta
\right) }-P_{\left( 2\right) }\right) $ . Furthermore, since (see also [8])

\begin{equation}
\int dy=\frac{1}{2\pi }\int d\sigma \left( q_{\left( \delta \right) }\right)
d\sigma \left( P_{\left( 2\right) }\right) \delta _{4}\left( q_{\left(
\gamma \right) }+P_{\left( 1\right) }-q_{\left( \delta \right) }-P_{\left(
2\right) }\right) ,d\sigma \left( q\right) =\left( \frac{d^{3}q}{q^{0}}%
\right)  \tag{16}
\end{equation}%
the normalized neutrino (or charged lepton) energy transfer \ $y=\left(
q_{\left( \gamma \right) }^{0}-q_{\left( \delta \right) }^{0}\right) \diagup
q_{\left( \gamma \right) }^{0}$ \ cannot affect Lorentz invariance of any of
the differential cross-sections.

Also, in view of (11), the cross-section normalization factor is defined
with respect to the massless flavor neutrino momenta.

\[
B=\frac{1}{\left( 2\pi \right) ^{6}}\left\vert \left( P_{\left( 1\right)
}\cdot q_{\left( \gamma \right) }\right) ^{2}-P_{\left( 1\right)
}^{2}q_{\left( \gamma \right) }^{2}\right\vert ^{\frac{1}{2}}=\frac{1}{%
\left( 2\pi \right) ^{6}}\left\vert \left( P_{\left( 1\right) }\cdot
q_{\left( \gamma \right) }\right) \right\vert 
\]%
In explicit evaluations, one uses the following short-hand notations:

\begin{equation}
m_{\alpha \beta }=\sum_{i}U_{\alpha i}m_{i}U_{i\beta }^{\dagger },m_{\alpha
\beta }^{2}=\sum_{i}U_{\alpha i}m_{i}^{2}U_{i\beta }^{\dagger }\text{ \ \ \
\ \ \ \ \ \ \ \ \ \ \ \ }  \tag{17}
\end{equation}

\bigskip

\textbf{\ Deriving the differential cross-sections with new energy
projection operators for the flavor neutrino processes within the massive
neutrino Fock space----}

\bigskip

$\frac{d\sigma _{W}}{dy}-$From the Lagrangian density in (4), the free
neutrino field (5), the kinematical relation (7), the relations (8) and (9),
one derives in the usual way the \ $W-$exchange $S_{W}$ and $S_{W}^{\dagger
} $ matrix elements for the process in (15). Specifically, with the Fierz
rearrangement and repeated indices summing up, one has,

\begin{eqnarray}
S_{W} &=&\sum_{i,j}\delta _{4}\left( q_{\left( i,\gamma \right) }+P_{\left(
1\right) }-q_{\left( j,\delta \right) }-P_{\left( 2\right) }\right) \delta
_{\alpha \beta }U_{\delta j}U_{i\gamma }^{\dagger }U_{j\beta }^{\dagger
}U_{\alpha i}\frac{ig^{2}}{\left( 2\pi \right) ^{2}8M_{W}^{2}}  \nonumber \\
&&\times \overline{u}\left( q_{\left( j,\delta \right) },s_{\left( j,\delta
\right) }\right) \gamma ^{\mu }\left( 1-\gamma ^{5}\right) u\left( q_{\left(
i,\gamma \right) },s_{\left( i,\gamma \right) }\right) \overline{u}\left(
P_{\left( 2\right) },r_{2}\right) \gamma _{\mu }u\left( P_{\left( 1\right)
,}r_{1}\right) \text{\ \ \ \ \ \ \ \ \ \ \ \ \ \ \ }  \TCItag{18}
\end{eqnarray}%
and $\ S_{W}^{\dagger }$ \ \ is obtained from (18) as shown in (10). The
contribution to the process (15) due to the $W-$exchange from (18), after
taking into account (11), (17), $\sqrt{2}g^{2}=8M_{W}^{2}G$, and the fact
that \ $s_{\left( i,\gamma \right) }$, $s_{\left( j,\delta \right) }$, ...,
obey, respectively, the ordinary and spinorial helicity relation, (8) and
(13), the standard procedure gives,

\begin{eqnarray}
s_{\left( i,\gamma \right) } &=&s_{\left( g,\gamma \right) }=s_{\left(
\gamma \right) };s_{\left( h,\delta \right) }=s_{\left( j,\delta \right)
}=s_{\left( \delta \right) }:  \nonumber \\
\frac{d\sigma _{W}\left( m\right) }{dy} &=&\frac{d\sigma _{W}^{\left(
c,c\right) }\left( m\right) }{dy}=\frac{G^{2}}{4\pi \left\vert \left(
P_{\left( 1\right) }\cdot q_{\left( \gamma \right) }\right) \right\vert }%
\frac{\delta _{\alpha \beta }}{2^{5}}\sum_{i,j;g,h}\left( U_{h\delta
}^{\dagger }U_{\gamma g}U_{\beta h}U_{g\alpha }^{\dagger }\right) \left(
U_{\delta j}U_{i\gamma }^{\dagger }U_{j\beta }^{\dagger }U_{\alpha i}\right)
\nonumber \\
&&\times \left[ Tr\left( M_{1}-\underline{P}_{\left( 1\right) }\right)
\gamma _{\nu }\left( 1-\gamma ^{5}\right) \left( M_{2}-\underline{P}_{\left(
2\right) }\right) \gamma _{\mu }\left( 1-\gamma ^{5}\right) \right] 
\TCItag{19} \\
&&\times \left[ Tr\left[ q_{\left( i,\gamma \right) },q_{\left( g,\gamma
\right) };+,c\right] \gamma ^{\nu }\left( 1-\gamma ^{5}\right) \left[
q_{\left( h,\delta \right) },q_{\left( j,\delta \right) };+,c\right] \gamma
^{\mu }\left( 1-\gamma ^{5}\right) \right]  \nonumber
\end{eqnarray}%
where $m$ symbolically denotes dependence on $m_{1,2,3\text{ . }}$Next, the
coherent energy operator expansion according to (14), with gamma matrices
traces carried out, yields

\begin{eqnarray}
\frac{d\sigma _{W}\left( m\right) }{dy} &=&\frac{d\sigma _{W}\left(
SM\right) }{dy}\left[ 1+\frac{m_{\alpha \alpha }^{2}}{4}\left( \frac{1}{%
q_{\left( \gamma \right) }^{02}}+\frac{1}{q_{\left( \delta \right) }^{02}}%
\right) \right] -\frac{2G^{2}\delta _{\alpha \beta }}{\pi \left\vert \left(
P_{\left( 1\right) }\cdot q_{\left( \gamma \right) }\right) \right\vert } 
\nonumber \\
&&\times \left\{ \frac{\delta _{\alpha \gamma }}{4}m_{\alpha \delta
}m_{\delta \alpha }\left[ \frac{\left( P_{\left( 1\right) }\cdot q_{\left(
\gamma \right) }\right) \left( P_{\left( 2\right) }\cdot q_{\left( \delta
\right) }\right) }{q_{\left( \delta \right) }^{02}}+\frac{2P_{\left(
2\right) }^{0}\left( P_{\left( 1\right) }\cdot q_{\left( \gamma \right)
}\right) }{q_{\left( \delta \right) }^{0}}\right] \right.  \TCItag{20} \\
&&\left. +\frac{\delta _{\beta \delta }}{4}m_{\alpha \gamma }m_{\gamma
\alpha }\left[ \frac{\left( P_{\left( 1\right) }\cdot q_{\left( \gamma
\right) }\right) \left( P_{\left( 2\right) }\cdot q_{\left( \delta \right)
}\right) }{q_{\left( \gamma \right) }^{02}}+\frac{2P_{\left( 1\right)
}^{0}\left( P_{\left( 2\right) }\cdot q_{\left( \delta \right) }\right) }{%
q_{\left( \gamma \right) }^{0}}\right] \right\} +O(m^{4}),  \nonumber \\
\frac{d\sigma _{W}\left( SM\right) }{dy} &=&\frac{2G^{2}\delta _{\alpha
\beta }\delta _{\alpha \gamma }\delta _{\beta \delta }}{\pi \left\vert
\left( P_{\left( 1\right) }\cdot q_{\left( \gamma \right) }\right)
\right\vert }\left( P_{\left( 1\right) }\cdot q_{\left( \gamma \right)
}\right) \left( P_{\left( 2\right) }\cdot q_{\left( \delta \right) }\right) 
\nonumber
\end{eqnarray}%
One can notice that , while the negligible $LIV$ is associated with the
neutrino mass, the $LI$ Standard Model result is formally identified with
zero neutrino mass limits

\begin{equation}
\frac{d\sigma _{W}\left( m\right) }{dy}=\frac{d\sigma _{W}\left( SM\right) }{%
dy}+O\left( m^{2};LIV\right)  \tag{21}
\end{equation}%
one can summarize the neutrino flavor transitions for the $W-$exchange
neutrino processes. Flavor conserving are: $LI$ to $O(m=0)$ terms and
negligible $LIV$ to $O(m^{2})$ terms. Flavor violating are: negligible $LIV$
to $O(m^{2})$ terms.

\bigskip

$\frac{d\sigma _{Z}}{dy}-$As in the previous case, from the Lagrangian
density in (4), the free neutrino field (5), the kinematical relation (7),
the contractions (9), one derives in the usual way the $Z-exchange$ $S_{Z}$
and $S_{Z}^{\dagger }$ matrix elements for the process in (15).
Specifically, one has

\begin{eqnarray}
S_{Z} &=&\sum_{i,j}\delta _{4}\left( q_{\left( i,\gamma \right) }+P_{\left(
1\right) }-q_{\left( j,\delta \right) }-P_{\left( 2\right) }\right) \delta
_{ij}\delta _{\alpha \beta }U_{\delta j}U_{i\gamma }^{\dagger }  \nonumber \\
&&\times \frac{ig^{2}}{\left( 2\pi \right) ^{2}16c_{W}^{2}M_{Z}^{2}}\left[ 
\overline{u}\left( q_{\left( j,\delta \right) },s_{\left( j,\delta \right)
}\right) \gamma ^{\mu }\left( 1-\gamma ^{5}\right) u\left( q_{\left(
i,\gamma \right) },s_{\left( i,\gamma \right) }\right) \right.  \TCItag{22}
\\
&&\left. \times \overline{u}\left( P_{\left( 2\right) },r_{2}\right) \gamma
_{\mu }\left( 4s_{W}^{2}-1+\gamma ^{5}\right) u\left( P_{\left( 1\right)
,}r_{1}\right) \right]  \nonumber
\end{eqnarray}%
while \ $S_{Z}^{\dagger }$ is obtained from the $S_{Z}$ through the
Hermitian conjugation. \ In what follows, one will find the following
shorthand notation very useful:

\begin{eqnarray}
w_{0} &=&s_{W}^{2},w_{1}=2s_{W}^{2}-1,z_{1}=s_{W}^{2}\left(
2s_{W}^{2}-1\right) +\frac{1}{4},z_{2}=s_{W}^{2}\left( 2s_{W}^{2}-1\right) ,
\nonumber \\
z_{3} &=&s_{W}^{2}-\frac{1}{4},z_{4}=s_{W}^{2}\left( s_{W}^{2}-1\right) +%
\frac{1}{4},z_{1}+z_{3}=2s_{W}^{4}  \TCItag{23}
\end{eqnarray}

After taking into account that $c_{W}^{2}M_{Z}^{2}=M_{W}^{2}$ and the fact
that the helicities, $s_{\left( i,\gamma \right) },s_{\left( j,\delta
\right) },...$, obey both the ordinary and the spinorial helicity relations
(8) and (13), the standard procedure yields the general expressions:

\begin{eqnarray}
s_{\left( i,\gamma \right) } &=&s_{\left( k,\gamma \right) }=s_{\left(
\gamma \right) };s_{\left( l,\delta \right) }=s_{\left( j,\delta \right)
}=s_{\left( \delta \right) }:  \nonumber \\
\frac{d\sigma _{Z}\left( m\right) }{dy} &=&\frac{d\sigma _{Z}^{\left(
c,c\right) }\left( m\right) }{dy}=\frac{G^{2}}{4\pi \left\vert \left(
P_{\left( 1\right) \cdot }q_{\left( \gamma \right) }\right) \right\vert }%
\frac{\delta _{\alpha \beta }}{2^{5}}\sum_{i,j;k,l}\left( U_{\delta
j}U_{i\gamma }^{\dagger }U_{l\delta }^{\dagger }U_{\gamma k}\right) \delta
_{ij}\delta _{kl}  \TCItag{24} \\
&&\times \left[ Tr\left( M_{1}-\underline{P}_{\left( 1\right) }\right)
\gamma _{\nu }\left( 2z_{3}+\frac{1}{2}\gamma ^{5}\right) \left( M_{2}-%
\underline{P}_{\left( 2\right) }\right) \gamma _{\mu }\left( 2z_{3}+\frac{1}{%
2}\gamma ^{5}\right) \right]  \nonumber \\
&&\times \left[ Tr\left[ q_{\left( i,\gamma \right) },q_{\left( k,\gamma
\right) };+,c\right] \gamma ^{\nu }\left( 1-\gamma ^{5}\right) \left[
q_{\left( l,\delta \right) },q_{\left( j,\delta \right) };+,c\right] \gamma
^{\mu }\left( 1-\gamma ^{5}\right) \right]  \nonumber
\end{eqnarray}%
The coherent energy operator expansion according to (14), with evaluating
the traces of gamma matrices, yields

\begin{eqnarray}
\frac{d\sigma _{Z}\left( m\right) }{dy} &=&\frac{d\sigma _{Z}\left(
SM\right) }{dy}\left[ 1+\frac{m_{\gamma \gamma }^{2}}{4}\left( \frac{1}{%
q_{\left( \gamma \right) }^{02}}+\frac{1}{q_{\left( \delta \right) }^{02}}%
\right) \right] -\frac{G^{2}m_{\gamma \delta }m_{\delta \gamma }\delta
_{\alpha \beta }}{8\pi \left\vert \left( P_{\left( 1\right) \cdot }q_{\left(
\gamma \right) }\right) \right\vert }  \nonumber \\
&&\times \left\{ 2\left( \frac{1}{q_{\left( \gamma \right) }^{02}}+\frac{1}{%
q_{\left( \delta \right) }^{02}}\right) \left[ M_{1}M_{2}z_{2}\left(
q_{\left( \gamma \right) }\cdot q_{\left( \delta \right) }\right) \right.
\right.  \nonumber \\
&&+\left( P_{\left( 1\right) }\cdot q_{\left( \delta \right) }\right) \left(
P_{\left( 2\right) }\cdot q_{\left( \gamma \right) }\right) \left(
z_{1}+z_{3}\right)  \nonumber \\
&&\left. +\left( P_{\left( 1\right) }\cdot q_{\left( \gamma \right) }\right)
\left( P_{\left( 2\right) }\cdot q_{\left( \delta \right) }\right) \left(
z_{1}-z_{3}\right) \right]  \nonumber \\
&&+\frac{4}{q_{\left( \delta \right) }^{0}}\left[ M_{1}M_{2}z_{2}q_{\left(
\gamma \right) }^{0}+P_{\left( 1\right) }^{0}\left( P_{\left( 2\right)
}\cdot q_{\left( \gamma \right) }\right) \left( z_{1}+z_{3}\right) \right. 
\TCItag{25} \\
&&\left. +P_{\left( 2\right) }^{0}\left( P_{\left( 1\right) }\cdot q_{\left(
\gamma \right) }\right) \left( z_{1}-z_{3}\right) \right] +\frac{4}{%
q_{\left( \gamma \right) }^{0}}\left[ M_{1}M_{2}z_{2}q_{\left( \delta
\right) }^{0}\right.  \nonumber \\
&&\left. \left. +P_{\left( 2\right) }^{0}\left( P_{\left( 1\right) }\cdot
q_{\left( \delta \right) }\right) \left( z_{1}+z_{3}\right) +P_{\left(
1\right) }^{0}\left( P_{\left( 2\right) }\cdot q_{\left( \delta \right)
}\right) \left( z_{1}-z_{3}\right) \right] \right\}  \nonumber \\
&&+O(m^{4})\text{\ }  \nonumber \\
\frac{d\sigma _{Z}\left( SM\right) }{dy} &=&\frac{G^{2}\delta _{\alpha \beta
}\delta _{\gamma \delta }}{\pi \left\vert \left( P_{\left( 1\right) \cdot
}q_{\left( \gamma \right) }\right) \right\vert }  \nonumber \\
&&\times \left[ M_{1}M_{2}z_{2}\left( q_{\left( \gamma \right) }\cdot
q_{\left( \delta \right) }\right) +\left( P_{\left( 1\right) }\cdot
q_{\left( \delta \right) }\right) \left( P_{\left( 2\right) }\cdot q_{\left(
\gamma \right) }\right) \left( z_{1}+z_{3}\right) \right. \text{ }  \nonumber
\\
&&\left. +\left( P_{\left( 1\right) }\cdot q_{\left( \gamma \right) }\right)
\left( P_{\left( 2\right) }\cdot q_{\left( \delta \right) }\right) \left(
z_{1}-z_{3}\right) \right] \text{\ \ \ \ \ \ \ \ \ \ \ \ \ \ \ \ \ \ \ \ \ \
\ \ \ \ \ \ \ \ \ \ \ \ \ \ \ \ \ \ \ \ \ \ \ \ \ \ \ \ \ \ \ \ \ \ \ \ \ \
\ \ \ \ \ \ \ \ \ \ \ \ }  \nonumber
\end{eqnarray}%
Here also,the negligible $LIV$ is associated with the neutrino mass while
the $LI$ Standard Model result is identified with formally zero neutrino
mass limits :

\begin{equation}
\frac{d\sigma _{Z}\left( m\right) }{dy}=\frac{d\sigma _{Z}\left( SM\right) }{%
dy}+O\left( m^{2};LIV\right) ,  \tag{26}
\end{equation}%
While the terms of $O\left( m=0\right) $ are $LI$ \ and flavor conserving,
the negligible $LIV$ terms \ of $O(m^{2})$are either flavor violating or
flavor conserving.

\bigskip

$\frac{d\sigma _{\left\{ W,Z\right\} }}{dy}-$Here, the differential
cross-section for the quasi-elastic neutrino scattering (15) due to the
overlapping $S-matrix$ elements from the $W-$ and $Z-$is given as a sum of
its components after taking into account relations (11), (18), and (22).
Importantly, again taking into account the fact that helicities, $s_{\left(
i,\gamma \right) ,s_{\left( j,\delta \right) }},...,$ obey both the ordinary
and spinorial helicity relations, (8) and (13), the standard procedure
yields the general expression

\begin{eqnarray}
s_{\left( e,\gamma \right) } &=&s_{\left( g,\gamma \right) }=s_{\left(
i,\gamma \right) }=s_{\left( k,\gamma \right) }=s_{\left( \gamma \right) }; 
\nonumber \\
s_{\left( h,\delta \right) } &=&s_{\left( f,\delta \right) }=s_{\left(
l,\delta \right) }=s_{\left( j,\delta \right) }=s_{\left( \delta \right) }: 
\nonumber \\
\frac{d\sigma _{\left\{ W,Z\right\} }\left( m\right) }{dy} &=&\frac{d\sigma
_{\left\{ W,Z\right\} }^{\left( c,c\right) }\left( m\right) }{dy}=\frac{G^{2}%
}{8\pi \left\vert \left( P_{\left( 1\right) \cdot }q_{\left( \gamma \right)
}\right) \right\vert }\frac{\delta _{\alpha \beta }}{2^{5}}  \nonumber \\
&&\times \left\{ \left[ Tr\left( M_{1}-\underline{P}_{\left( 1\right)
}\right) \gamma _{\nu }\left( 4z_{3}+\gamma ^{5}\right) \left( M_{2}-%
\underline{P}_{\left( 2\right) }\right) \gamma _{\mu }\left( 1-\gamma
^{5}\right) \right] \right.  \nonumber \\
&&\times \left\lfloor \left[ \sum_{g,h;e,f}\left( U_{h\delta }^{\dagger
}U_{\gamma g}U_{\beta h}U_{g\alpha }^{\dagger }U_{\delta f}U_{e\gamma
}^{\dagger }\delta _{ef}\right) \right. \right.  \TCItag{27} \\
&&\left. \times Tr\left[ q_{\left( e,\gamma \right) },q_{\left( g,\gamma
\right) };+,c\right] \gamma ^{\nu }\left( 1-\gamma ^{5}\right) \left[
q_{\left( h,\delta \right) },q_{\left( f,\delta \right) };+,c\right] \gamma
^{\mu }\left( 1-\gamma ^{5}\right) \right]  \nonumber \\
&&+\left[ \sum_{k,l;i,j}\left( U_{l\delta }^{\dagger }U_{\gamma k}U_{\delta
j}U_{i\gamma }^{\dagger }U_{\alpha i}U_{j\beta }^{\dagger }\delta
_{kl}\right) \right.  \nonumber \\
&&\left. \left. \left. \times Tr\left[ q_{\left( i,\gamma \right)
},q_{\left( k,\gamma \right) };+,c\right] \gamma ^{\nu }\left( 1-\gamma
^{5}\right) \left[ q_{\left( l,\delta \right) },q_{\left( j,\delta \right)
};+,c\right] \gamma ^{\mu }\left( 1-\gamma ^{5}\right) \right] \right\rfloor
\right\}  \nonumber
\end{eqnarray}%
where one took into account the identity:%
\begin{eqnarray*}
&&Tr\left( M_{1}-\underline{P}_{\left( 1\right) }\right) \gamma _{\nu
}\left( 4z_{3}+\gamma ^{5}\right) \left( M_{2}-\underline{P}_{\left(
2\right) }\right) \gamma _{\mu }\left( 1-\gamma ^{5}\right) \\
&=&Tr\left( M_{1}-\underline{P}_{\left( 1\right) }\right) \gamma _{\nu
}\left( 1-\gamma ^{5}\right) \left( M_{2}-\underline{P}_{\left( 2\right)
}\right) \gamma _{\mu }\left( 4z_{3}+\gamma ^{5}\right)
\end{eqnarray*}%
Of course, one cannot avoid the coherent energy operator expansion according
to (14), and evaluating the traces of gamma matrices one obtains

\begin{eqnarray}
\frac{d\sigma _{\left\{ W,Z\right\} }\left( m\right) }{dy} &=&\frac{d\sigma
_{\left\{ W,Z\right\} }\left( SM\right) }{dy}\left[ 1+\frac{m_{\gamma \gamma
}^{2}}{4}\left( \frac{1}{q_{\left( \gamma \right) }^{02}}+\frac{1}{q_{\left(
\delta \right) }^{02}}\right) \right] -\frac{G^{2}m_{\gamma \delta
}m_{\delta \gamma }\delta _{\alpha \beta }}{2\pi \left\vert \left( P_{\left(
1\right) \cdot }q_{\left( \gamma \right) }\right) \right\vert }  \nonumber \\
&&\times \left\{ \left[ M_{1}M_{2}w_{0}\left( q_{\left( \gamma \right)
}\cdot q_{\left( \delta \right) }\right) +w_{1}\left( P_{\left( 1\right)
\cdot }q_{\left( \gamma \right) }\right) \left( P_{\left( 2\right) \cdot
}q_{\left( \delta \right) }\right) \right] \right.  \nonumber \\
&&\times \left( \frac{\delta _{\alpha \gamma }}{q_{\left( \gamma \right)
}^{02}}+\frac{\delta _{\beta \delta }}{q_{\left( \delta \right) }^{02}}%
\right) +\frac{2\delta _{\alpha \gamma }}{q_{\left( \delta \right) }^{0}}%
\left[ M_{1}M_{2}w_{0}q_{\left( \gamma \right) }^{0}+w_{1}\left( P_{\left(
1\right) \cdot }q_{\left( \gamma \right) }\right) P_{\left( 2\right) }^{0}%
\right]  \TCItag{28} \\
&&\left. +\frac{2\delta _{\beta \delta }}{q_{\left( \gamma \right) }^{0}}%
\left[ M_{1}M_{2}w_{0}q_{\left( \delta \right) }^{0}+w_{1}\left( P_{\left(
2\right) \cdot }q_{\left( \delta \right) }\right) P_{\left( 1\right) }^{0}%
\right] \right\} +O\left( m^{4}\right) ,  \nonumber \\
\frac{d\sigma _{\left\{ W,Z\right\} }\left( SM\right) }{dy} &=&\frac{%
2G^{2}\delta _{\alpha \beta }\delta _{\alpha \gamma }\delta _{\gamma \delta }%
}{\pi \left\vert \left( P_{\left( 1\right) \cdot }q_{\left( \gamma \right)
}\right) \right\vert }\left[ M_{1}M_{2}w_{0}\left( q_{\left( \gamma \right)
}\cdot q_{\left( \delta \right) }\right) +w_{1}\left( P_{\left( 1\right)
\cdot }q_{\left( \gamma \right) }\right) \left( P_{\left( 2\right) \cdot
}q_{\left( \delta \right) }\right) \right] \text{\ \ \ \ \ \ \ \ }  \nonumber
\end{eqnarray}%
Again,the negligible $LIV$ is associated with the neutrino mass while the $%
LI $ Standard Model result is identified with formally zero neutrino mass
limits :

\begin{equation}
\frac{d\sigma _{\left\{ W,Z\right\} }\left( m\right) }{dy}=\frac{d\sigma
_{\left\{ W,Z\right\} }\left( SM\right) }{dy}+O\left( m^{2};LIV\right) 
\tag{29}
\end{equation}%
The overlapping $W-$and \ $Z-$ exchanges cross-section terms of $O\left(
m=0\right) $ are $LI$ \ and flavor conserving, while the negligible terms of 
$O(m^{2})$ carry the $LIV$ terms with, both, the conserved and violated
flavor.

\bigskip

\textbf{\ Phenomenological neutrino cross-section intensity from time
extrapolation of the neutrino oscillation scattering }

\bigskip

Dvornikov in the classical field theoretical model [17] quotes the neutrino
oscillation-transition probability with the characteristic Pontecorvo
dimensionless argument $\ \Delta m^{2}t/4E$ ,

\begin{equation}
P\left( t\right) =\sin ^{2}\left( 2\Theta _{Vac}\right) \sin \left( \frac{%
\Delta m^{2}t}{4E}\right)  \tag{30}
\end{equation}%
where $\Theta _{Vac}$ is the vaccum mixing angle, $\Delta
m^{2}=m_{1}^{2}-m_{2}^{2}$ is the mass squared difference and $E$ is the
energy of the system (detailed description of these and other parameters is
in [17]). The interest in [17] comes from the fact that at $t=E^{-1}$ one
can expand (3) for small Pontecorvo argument as sin $\left( \frac{\Delta
m^{2}}{4E^{2}}\right) =\frac{\Delta m^{2}}{4E^{2}}+....$. Or, turning the
logic around, one could think of the scattering theory presented here as an
approximation of a more general description, yet to be derived, where sin $%
\left( \frac{\Delta m^{2}}{4E^{2}}\right) $ \ occurred \ but for the
laboratory measurments was approximated with $\frac{\Delta m^{2}}{4E^{2}}$.
So, with this reversed logic, one can assume that the calculated neutrino
intensity , defined appropriately from differential cross-sections, can be
analyzed at the practical baseline region if in it every dimensionless
Pontecorvo argument of the type $\frac{\Delta m^{2}}{4E^{2}}$ is replaced
with $\sin \left( \frac{\Delta m^{2}t}{4E}\right) $. Other terms remain
unaffected and, if proportional to the squares of neutrino masses, may even
be neglected. However, the recoil, which Dvornikov [17] in the classical
approach did not have , here will have to be taken into account and, if
possible, averaged out out in the calculated neutrino intensity in the short
baseline region .

To pursue the idea from reference [17] on the phenomenological level, say,
on the neutrino scattering process as described by the differential
cross-section (28), it is necessary to work in the laboratory frame, $%
P_{\left( 1\right) }=\left( \overrightarrow{0},M_{1}\right) $ . Next thing
is to concentrate on the incoming neutrino energy in (28) \ with the help of
the neutrino energy transfer

\begin{eqnarray}
y &=&\frac{q_{\left( \gamma \right) }^{0}-q_{\left( \delta \right) }^{0}}{%
q_{\left( \gamma \right) }^{0}}=\frac{P_{\left( 2\right) }^{0}-M_{1}}{%
q_{\left( \gamma \right) }^{0}}  \TCItag{31} \\
q_{\left( \delta \right) }^{0} &=&\left( 1-y\right) q_{\left( \gamma \right)
}^{0}  \nonumber
\end{eqnarray}%
So that after substituting (31) into (28) one obtains for the differential
cross-section in (28)

\begin{eqnarray}
\frac{d\sigma _{\left\{ W,Z\right\} }\left( m\right) }{dy} &=&\frac{d%
\overline{\sigma }_{\left\{ W,Z\right\} }\left( SM\right) }{dy}\delta
_{\alpha \beta }\left\{ \delta _{\alpha \gamma }\delta _{\gamma \delta
}+\delta _{\alpha \gamma }\delta _{\gamma \delta }\frac{m_{\gamma \gamma
}^{2}}{4q_{\left( \gamma \right) }^{02}}\left[ 1+\frac{1}{(1-y)^{2}}\right]
\right.  \nonumber \\
&&\left. -\frac{m_{\gamma \delta }m_{\delta \gamma }}{4q_{\left( \gamma
\right) }^{02}}\left[ \delta _{\alpha \gamma }+\frac{\delta _{\beta \delta }%
}{1-y)^{2}}\right] \right\}  \TCItag{32} \\
&&-\frac{G^{2}m_{\gamma \delta }m_{\delta \gamma }\delta _{\alpha \beta }}{%
\pi \left\vert \left( P_{\left( 1\right) \cdot }q_{\left( \gamma \right)
}\right) \right\vert }\left\{ w_{0}M_{1}M_{2}\left( \frac{\delta _{\alpha
\gamma }}{\left( 1-y\right) }+\delta _{\beta \delta }\left( 1-y\right)
\right) \right.  \nonumber \\
&&\left. +w_{1}\left( -\frac{\delta _{\alpha \gamma }M_{1}P_{\left( 2\right)
}^{0}}{\left( 1-y\right) }+\frac{\delta _{\beta \delta }\left( P_{\left(
2\right) \cdot }q_{\left( \delta \right) }\right) M_{1}}{q_{\left( \gamma
\right) }^{0}}\right) \right\}  \nonumber
\end{eqnarray}

\begin{eqnarray}
\frac{d\overline{\sigma }_{\left\{ W,Z\right\} }\left( SM\right) }{dy} &=&%
\frac{2G^{2}}{\pi \left\vert \left( P_{\left( 1\right) \cdot }q_{\left(
\gamma \right) }\right) \right\vert }\left[ M_{1}M_{2}w_{0}\left( q_{\left(
\gamma \right) }\cdot q_{\left( \delta \right) }\right) \right.  \nonumber \\
&&\left. -w_{1}M_{1}q_{\left( \gamma \right) }^{0}\left( P_{\left( 2\right)
\cdot }q_{\left( \delta \right) }\right) \right]  \TCItag{33}
\end{eqnarray}

In order to bring out the new features, one "removes" he SM cross-section
from the newly derived cross-section, through the definition of the neutrino
cross-section intensity which, as suggested by (30), will allow the
introduction of the dimensionless Pontecorvo argument (30).

\begin{eqnarray}
I_{\left\{ W,Z\right\} }\left( m,\theta \right) &=&\left[ \frac{d\sigma
_{\left\{ W,Z\right\} }\left( m\right) }{dy}-\frac{d\overline{\sigma }%
_{\left\{ W,Z\right\} }\left( SM\right) }{dy}\right] \left[ \frac{d\overline{%
\sigma }_{\left\{ W,Z\right\} }\left( SM\right) }{dy}\right] ^{-1}  \nonumber
\\
&=&\delta _{\alpha \beta }\left\{ \delta _{\alpha \gamma }\delta _{\gamma
\delta }\frac{m_{\gamma \gamma }^{2}}{4q_{\left( \gamma \right) }^{02}}\left[
1+\frac{1}{(1-y)^{2}}\right] \right.  \nonumber \\
&&\left. -\frac{m_{\gamma \delta }m_{\delta \gamma }}{4q_{\left( \gamma
\right) }^{02}}\left[ \delta _{\alpha \gamma }+\frac{\delta _{\beta \delta }%
}{1-y)^{2}}\right] \right\}  \TCItag{34} \\
&&-\frac{m_{\gamma \delta }m_{\delta \gamma \delta _{\alpha \beta }}}{2}%
\left\{ \left[ w_{0}M_{1}M_{2}\left( \frac{\delta _{\alpha \gamma }}{(1-y)}%
+\delta _{\beta \delta }(1-y)\right) \right. \right.  \nonumber \\
&&\left. +w_{1}\left( -\frac{\delta _{\alpha \gamma }M_{1}P_{\left( 2\right)
}^{0}}{(1-y)}+\frac{\delta _{\beta \delta }\left( P_{\left( 2\right) \cdot
}q_{\left( \delta \right) }\right) M_{1}}{q_{\left( \gamma \right) }^{0}}%
\right) \right]  \nonumber \\
&&\left. \times \left[ w_{0}M_{1}M_{20}\left( q_{\left( \gamma \right)
}\cdot q_{\left( \delta \right) }\right) -w_{1}M_{1}q_{\left( \gamma \right)
}^{0}\left( P_{\left( 2\right) \cdot }q_{\left( \delta \right) }\right) %
\right] ^{-1}\right\}  \nonumber
\end{eqnarray}%
\begin{equation}
\overrightarrow{q}_{\left( \gamma \right) }\cdot \overrightarrow{P}_{\left(
2\right) }=q_{\left( \gamma \right) }^{0}\left\vert \overrightarrow{P}%
_{\left( 2\right) }\right\vert \cos \theta ,\text{ }y=\frac{2M_{1}q_{\left(
\gamma \right) }^{0}\cos ^{2}\theta }{\left( M_{1}+q_{\left( \gamma \right)
}^{0}\right) ^{2}-q_{\left( \gamma \right) }^{02}\cos ^{2}\theta }  \tag{35}
\end{equation}%
Where the dimensionless intensity $I_{\left\{ W,Z\right\} }\left( m,\theta
\right) $ from (34) will serve also as unnormalized probability. As shown in
(31) and (35), the angle $\theta $-dependence comes through the kinematics.
The expression for the normalized neutrino energy transfer $y$ \ follows
from scattering kinematics and the energy-momentum conservation (more
details can be found also in [9]). As seen in (35), for fixed $q_{\left(
\gamma \right) }^{0}$, $y$ depends only on \ $\cos ^{2}\theta $ which could
be averaged over straightforwardly.

The relation (34), of course, is an instantaneous intensity at the time of
interaction. This instant can be defined as $\ t_{0}=1/q_{\left( \gamma
\right) }^{0}$. For $q_{\left( \gamma \right) }^{0}=10MeV,$ $%
t_{0}=6.58\times 10^{-23}\sec .$;one can also define a "distance" \ with $%
L_{0}=1/q_{\left( \gamma \right) }^{0}$ to give for the same neutrino energy
\ $L_{0}=1.97\times 10^{-12}cm$. Of course, what counts are relative times
and distances. Here, $t_{0}$ and $L_{0}$ are given for the sake of
convenience. \ Now, as is described at the beginning of this section, one
can perform the time extrapolation with sinus functions on the arguments
that are in the Pontecorvo \ forms while others, with neutrino mass square
dependences, can be dropped. The result of this procedure is, where for the
sake of simplicity, the processes are denoted with just the flavor quantum
numbers. One has

\begin{eqnarray}
I_{\left\{ W,Z\right\} }\left( m,\theta \right) &=&I_{1}\left( P_{1\left(
\alpha \right) }+\nu _{\left( \gamma \right) }\rightarrow P_{2\left( \beta
\right) }+\nu _{\left( \delta \right) }\right)  \nonumber \\
&&+I_{2}\left( P_{1\left( \alpha \right) }+\nu _{\left( \gamma \right)
}\rightarrow P_{2\left( \beta \right) }+\nu _{\left( \delta \right) }\right)
\nonumber \\
&&+I_{3}\left( P_{1\left( \alpha \right) }+\nu _{\left( \gamma \right)
}\rightarrow P_{2\left( \beta \right) }+\nu _{\left( \delta \right) }\right)
,  \nonumber \\
I_{1}\left( P_{1\left( \alpha \right) }+\nu _{\left( \gamma \right)
}\rightarrow P_{2\left( \beta \right) }+\nu _{\left( \delta \right) }\right)
&=&\delta _{\alpha \beta }\delta _{\alpha \gamma }\delta _{\gamma \delta } 
\left[ 1+\frac{1}{(1-y)^{2}}\right] \sin \frac{m_{\gamma \gamma }^{2}t}{%
4q_{\left( \gamma \right) }^{0}},  \nonumber \\
I_{2}\left( P_{1\left( \alpha \right) }+\nu _{\left( \gamma \right)
}\rightarrow P_{2\left( \beta \right) }+\nu _{\left( \delta \right) }\right)
&=&-\delta _{\alpha \beta }\delta _{\alpha \gamma }\sin \frac{m_{\gamma
\delta }m_{\delta \gamma }t}{4q_{\left( \gamma \right) }^{0}},  \TCItag{36}
\\
I_{3}\left( P_{1\left( \alpha \right) }+\nu _{\left( \gamma \right)
}\rightarrow P_{2\left( \beta \right) }+\nu _{\left( \delta \right) }\right)
&=&-\delta _{\alpha \beta }\delta _{\beta \delta }\frac{1}{(1-y)^{2}}\sin 
\frac{m_{\gamma \delta }m_{\delta \gamma }t}{4q_{\left( \gamma \right) }^{0}}%
;  \nonumber \\
\left\langle I_{\left\{ W,Z\right\} }\left( m\right) \right\rangle &=&\frac{2%
}{\pi }\int_{0}^{\pi }d\theta I_{\left\{ W,Z\right\} }\left( m,\theta
\right) =\sum\limits_{i=1}^{3}\left\langle I_{i}\right\rangle  \TCItag{37}
\end{eqnarray}%
One notices that, since $q_{\left( \gamma \right) }^{0}$ is the incoming
neutrino energy, the intensities in (36) depend on the scattering angle $%
\theta $ only through $y$ (compare with (35)). So in (37) it is shown how to
get averaged, over the scattering angle, the total intensity $\left\langle
I_{\left\{ W,Z\right\} }\left( m\right) \right\rangle $ from the sum of
individual averaged intensities. Since the dimensionless intensities serve
also as unnormalized probabilities, one has to look just at zero to positive
values when connecting to observations; this has to be done for the
individual as well as for the total intensities, unaveraged and averaged:

\begin{eqnarray}
Observ.. &:&I_{i}\geq 0,\text{ }\left\langle I_{i}\right\rangle \geq 0 
\nonumber \\
Unobserv. &:&I_{i}\leq 0,\left\langle I_{i}\right\rangle \leq 0  \TCItag{38}
\end{eqnarray}

\begin{eqnarray}
Observ. &:&I_{\left\{ W,Z\right\} }\left( m,\theta \right) \geq
0,I_{i}\gtrless 0;\left\langle I_{\left\{ W,Z\right\} }\left( m,\theta
\right) \right\rangle \geq 0,\left\langle I_{i}\right\rangle \gtrless 0 
\nonumber \\
Unobserv. &:&I_{\left\{ W,Z\right\} }\left( m,\theta \right) \leq
0,I_{i}\gtrless 0;\left\langle I_{\left\{ W,Z\right\} }\left( m,\theta
\right) \right\rangle \leq 0,\left\langle I_{i}\right\rangle \gtrless 0 
\TCItag{39}
\end{eqnarray}

One notices that (36) has three independent parts as the flavor numbers $%
\gamma $ and $\delta $ are generally independent from $\alpha $ and $\beta $%
. Hence (36) is describing, respectively, the flavor conserving $\alpha
=\beta =\gamma =\delta $ transition, and the flavor violating $\ \alpha
=\beta =\gamma \neq \delta $\ and $\alpha =\beta =\delta \neq \gamma $
transitions. Already at this point, one can see that one has here the
physics beyond the SM. The interesting thing is that if one forces $\gamma $
and $\delta $ to be identical with $\alpha $ one obtains that $%
(36),(37)\approx 0$ for any value of $t$ or $L$ ; this is another indication
that physics went beyond the SM. \ \ \ \ \ 

\bigskip At this point, it the easiest to continue if one has the explicit
values for neutrino masses and the neutrino mixing matrix. The masses
accepted here are from the analysis by Fritzsch [18] with values,%
\begin{equation}
m_{1}=0.004eV,\text{ }m_{2}=0.01eV,\text{ }m_{3}=0.05eV  \tag{40}
\end{equation}%
while the neutrino mixing matrix is due to Harrison, Perkins and Scott [19]

\begin{equation}
\left( U_{\alpha i}\right) =\left( 
\begin{array}{ccc}
\sqrt{\frac{2}{3}} & \sqrt{\frac{1}{3}} & 0 \\ 
-\sqrt{\frac{1}{6}} & \sqrt{\frac{1}{3}} & -\sqrt{\frac{1}{2}} \\ 
-\sqrt{\frac{1}{6}} & \sqrt{\frac{1}{3}} & \sqrt{\frac{1}{2}}%
\end{array}%
\right)  \tag{41}
\end{equation}%
Now, taking that the target at rest is an electron, one has $%
M_{1}=M_{2}\approx 0.5MeV$, while one retains the incoming energy of the
neutrino at $q_{\left( \gamma \right) }^{0}=10MeV$ \ and looks for $%
L_{+}=t_{+}c$, and $L_{-}=t_{-}c$ , at which the values of $+$sinus and $-$%
sinus functions in (36) become $+1$ , yielding first large intensities. The
result is 
\begin{equation}
\sin \frac{m_{\gamma \gamma }^{2}t_{+}}{4q_{\left( \gamma \right) }^{0}}%
=1\rightarrow L_{+}=\frac{2\pi q_{\left( \gamma \right) }^{0}}{m_{\gamma
\gamma }^{2}};-\sin \frac{m_{\gamma \delta }m_{\delta \gamma }t_{-}}{%
4q_{\left( \gamma \right) }^{0}}=1\rightarrow L_{-}=\frac{6\pi q_{\left(
\gamma \right) }^{0}}{m_{\gamma \delta }m_{\delta \gamma }}  \tag{42}
\end{equation}%
Next, one realizes that, to a good approximation, these distances away from
the source, where the first large intensities occur, are given \ as $\Delta
L_{\pm }=L_{\pm }-L_{0}\thickapprox L_{\pm }$, that is, with just $L_{\pm }$%
. As to the specific place where one notices these maxima, one has to
involve the kinematics from (31) and (35).

Relation (42) is a general relation; as already mentioned, from (36), one
actually has three possibilities for transitions: one flavor conserving and
two flavor violating. Using the values for the neutrino masses in (40) and
the neutrino matrix in (41), one deduces the distances from (42) to be

\begin{eqnarray}
&&(Flav.conserv.)I_{1}\left( P_{1\left( \alpha \right) }+\nu _{\left( \gamma
\right) }\rightarrow P_{2\left( \beta \right) }+\nu _{\left( \delta \right)
}\right) \delta _{\alpha \beta }\delta _{\alpha \gamma }\delta _{\gamma
\delta };  \nonumber \\
\alpha ,\beta &=&e;\gamma ,\delta =e:L_{+}\left( \nu _{\left( e\right)
}\rightarrow \nu _{\left( e\right) }\right) \approx 281km.  \nonumber \\
&&\left( Flav.viol.\right) I_{2}\left( P_{1\left( \alpha \right) }+\nu
_{\left( \gamma \right) }\rightarrow P_{2\left( \beta \right) }+\nu _{\left(
\delta \right) }\right) \delta _{\alpha \beta }\delta _{\alpha \gamma }; 
\nonumber \\
\alpha ,\beta ,\gamma &=&e;\delta =\mu :L_{-}\left( \nu _{\left( e\right)
}\rightarrow \nu _{\left( \mu \right) }\right) \approx 9279km.  \TCItag{43}
\\
&&\left( Flav.viol.\right) I_{3}\left( P_{1\left( \alpha \right) }+\nu
_{\left( \gamma \right) }\rightarrow P_{2\left( \beta \right) }+\nu _{\left(
\delta \right) }\right) \delta _{\alpha \beta }\delta _{\beta \delta }; 
\nonumber \\
\alpha ,\beta ,\delta &=&e,\gamma =\mu :L_{-}\left( \nu _{\left( \mu \right)
}\rightarrow \nu _{\left( e\right) }\right) \approx 9279km.  \nonumber
\end{eqnarray}%
The distances are rather large, particularly for the flavor violating
oscillations, $\nu _{\left( e\right) }\rightarrow \nu _{\left( \mu \right) }$
and $\nu _{\left( \mu \right) }\rightarrow \nu _{\left( e\right) }$, which
one can attribute to the small values of neutrino masses. One notices that
among three distances in (43) two are the same, $L_{-}\left( \nu _{\left(
e\right) }\rightarrow \nu _{\left( \mu \right) }\right) =L_{-}\left( \nu
_{\left( \mu \right) }\rightarrow \nu _{\left( e\right) }\right) $.

\bigskip For $I_{1\text{ }}$and $I_{3}$ intensities it helps to look at $%
\theta $ dependence of $(1-y)^{-2}$ in order to see at which angles they are
best detected. Here are some of the most significant values.%
\begin{eqnarray}
&&(1-y)^{-2}\left( \theta =0;0.1;0.4;0.5;\frac{\pi }{2}\right)  \TCItag{44}
\\
&=&1681;60.34;2.26;1.7;1\ \ \   \nonumber
\end{eqnarray}%
Indeed, $I_{1}$ and $I_{3}$ have rather large maxima in forward, $\theta =0$%
, direction, direction that should be favorable for detecting neutrino
oscillation.

One may as well look at the average values of intensities over the
scattering angle $\theta .$For this one needs the scattering angle average \
quantity $\ \left\langle (1-y)^{-2}\right\rangle $ in (36).For the electron
target, which is a very light target, one notices very large values in the
forward , $\theta =0$ ,direction and rather smooth thereafter up to $\theta
=\pi /2$. The large value in question of $1681$ is at \ $\theta =0$. This
comes basically from the smallness of the electron mass. \ Another meaning
of the large value at $\theta =0$ is that the recoil activities are strong
at these angles when the target is light, and weak in the perpendicular
directions. The straightforward numerical integration from more detailed
evaluation of $\ (1-y)^{-2}$ yields for the average values%
\begin{eqnarray}
\left\langle (1-y)^{-2}\right\rangle &=&\frac{2}{\pi }\int_{0}^{\pi
/2}d\theta 1-y)^{-2}\approx 59,  \nonumber \\
\left\langle I_{1}\left( P_{1\left( \alpha \right) }+\nu _{\left( \gamma
\right) }\rightarrow P_{2\left( \beta \right) }+\nu _{\left( \delta \right)
}\right) \right\rangle &=&\delta _{\alpha \beta }\delta _{\alpha \gamma
}\delta _{\gamma \delta }60\sin \frac{m_{\gamma \gamma }^{2}t}{4q_{\left(
\gamma \right) }^{0}}  \nonumber \\
\left\langle I_{2}\left( P_{1\left( \alpha \right) }+\nu _{\left( \gamma
\right) }\rightarrow P_{2\left( \beta \right) }+\nu _{\left( \delta \right)
}\right) \right\rangle &=&-\delta _{\alpha \beta }\delta _{\alpha \gamma
}\sin \frac{m_{\gamma \delta }m_{\delta \gamma }t}{4q_{\left( \gamma \right)
}^{0}}  \TCItag{45} \\
\left\langle I_{3}\left( P_{1\left( \alpha \right) }+\nu _{\left( \gamma
\right) }\rightarrow P_{2\left( \beta \right) }+\nu _{\left( \delta \right)
}\right) \right\rangle &=&-\delta _{\alpha \beta }\delta _{\beta \delta
}59\sin \frac{m_{\gamma \delta }m_{\delta \gamma }t}{4q_{\left( \gamma
\right) }^{0}}  \nonumber
\end{eqnarray}%
The numbers 60 ,1 and 59 have the meanings or relative strengths with
respect to each other of these dimensionless intensities.\ Then according to
definitions in (38) and (39), for flavor conserving part, $\left\langle
I_{\left\{ W,Z\right\} }\left( m,\theta \right) \right\rangle \approx 100\%$ 
$\left\langle I_{1}\left( \alpha +\alpha \rightarrow \alpha +\alpha \right)
\right\rangle $, where one identified $60=100\%$. The flavor violating parts 
$\left\langle I_{2}\right\rangle $ and $\left\langle I_{3}\right\rangle $
one may say that they split 100\% as $\left\langle I_{\left\{ W,Z\right\}
}\left( m,\theta \right) \right\rangle \approx 2\%$ $\left\langle
I_{2}\left( \alpha +\alpha \rightarrow \alpha +\delta \right) \right\rangle
+98\%$ $\left\langle I_{3}\left( \alpha +\gamma \rightarrow \alpha +\alpha
\right) \right\rangle .$

Now, unlike in the classical field theoretical model of Dvornikov [17], here
one has quantum field theory with the particle kinematics so that scattering
angle $\theta $ between the incoming neutrino and scattered electron come
into play. So far one was dealing with the incoming neutrino and derived the
intensities. However, at the distances of $281km$ and $9279km$ one can hope
to detect the outgoing neutrino energy $q_{\left( \delta \right) }^{0}$
while, at the same time measure the scattering angle $\theta $ of the
recoiled electron. From relations (31) and (35) one now correlates $%
q_{\left( \delta \right) }^{0}$, $q_{\left( \gamma \right) }^{0}$ and angle $%
\theta $ through the relation

\begin{eqnarray}
P_{\left( 1\right) } &=&\left( \overrightarrow{0},M_{1}\right) ,%
\overrightarrow{q}_{\left( \gamma \right) }\cdot \overrightarrow{q}_{\left(
\delta \right) }=q_{\left( \gamma \right) }^{0}q_{\left( \delta \right)
}^{0}\cos \phi :  \nonumber \\
&&q_{\left( \gamma \right) }^{02}\sin ^{2}\theta \left[ q_{\left( \gamma
\right) }^{0}-q_{\left( \delta \right) }^{0}+2M_{1}\right]  \nonumber \\
+q_{\left( \gamma \right) }^{0}M_{1}\left( M_{1}-2q_{\left( \delta \right)
}^{0}\right) -q_{\left( \delta \right) }^{0}M_{1}^{2} &=&0;  \TCItag{46} \\
q_{\left( \gamma \right) }^{0} &=&\left\vert \overrightarrow{P}_{\left(
2\right) }\right\vert \cos \theta +q_{\left( \delta \right) }^{0}\cos \phi ,
\TCItag{47} \\
\left\vert \overrightarrow{P}_{\left( 2\right) }\right\vert \sin \theta
&=&q_{\left( \delta \right) }^{0}\sin \phi  \TCItag{48}
\end{eqnarray}%
Relation (46), for example, associates the incoming neutrino energy with \ a
measured outgoing neutrino energy. Relations (47) and (48) correlate the
electron recoil with incoming and outgoing neutrino energies. Specifically,
with the knowledge of $\theta $ and $q_{\left( \gamma \right) }^{0}$ , one
can calculate the value for $q_{\left( \delta \right) }^{0}$ ; for instance
for $\theta \approx 0$ , (46) gives $q_{\left( \delta \right) }^{0}\approx
0.24MeV$. \ From (47), one has that $\ \left\vert \cos \phi \right\vert
=\left\vert q_{\left( \gamma \right) }^{0}-\left\vert \overrightarrow{P}%
_{\left( 2\right) }\right\vert \cos \theta \right\vert \left( q_{\left(
\delta \right) }^{0}\right) ^{-1}\leq 1$. \ From (47) and/or (48), one can
solve for two quantities, assuming that others are known. The presumed
oscillation of the $\delta $-neutrino is occurring along the line with the
angle $\phi $ with respect to $\overrightarrow{q}_{\left( \gamma \right) }$
and the corresponding intensity is determined by relations (45). Although,
by and large, one expects $q_{\left( \delta \right) }^{0}<\left\vert 
\overrightarrow{P}_{\left( 2\right) }\right\vert $ which implies $\sin
\theta <\sin \phi $, never the less, smaller angle $\theta $ would imply
also smaller angle $\phi $. Of course in addition to relations (46) and (47)
one can find other relations that can serve the purpose of studying the
neutrino oscillations from scattering experiments.

\ \ \ \ \ \ \ \ \ \ \ \ \ \ \ .

\textbf{\ Discussion---} One thing that one notices right a way is the fact
\ that while the $LIV$ is very real, because it is associated with the $%
O(m^{2})$ terms, it is negligible at least in the scattering-like
experiments. Therefore, the "mass-less" SM is basically $LI$ because the
neutrinos have masses that are $\ \leq 1eV.$ Because the $LIV$ terms are
proportional to $O(m^{2})$ they play no role in the laboratory experiments
where the SM dominates. \ As shown by relations (30)- (34), the
extrapolation of the negligible portion of the laboratory neutrino
oscillation scattering cross-sections into the practical baseline
oscillation differential cross-sections should \ make them \ now observable
at reasonable distances from the interaction region, particularly through
the help of the neutrino cross-section intensity (40). The most interesting
thing is that this extrapolated differential cross-section contains the
flavor conserving and two kinds of flavor violating parts.

\bigskip

\textbf{Acknowledgements---} I wish to thank Dr. Howard E. Brandt for
friendly discussions and help with the computer manipulations. Thanks are
due to the referee (anonymous) for pointing out to additional and necessary
clarifications which extended the manuscript beyond the laboratory
scattering to the baseline oscillation region. To my wife Patricia Marie
Stone Soln, I am deeply grateful for her expert librarian help in locating
literature over the internet.

\bigskip

\textbf{\ References---}

[1]\qquad Super-KamiokandeCollaboration, Y. Ashie et al., Phys. Rev. Lett. 
\textbf{93}, 101801 (2004); M. Shiozawa, Prog. Part. Nucl. Phys. \textbf{57}%
, 79 (2006).

[2]\qquad SNO Collaboration, Phys. Rev. Lett. \textbf{81,} 071301 (2001); 
\textbf{89}, 011301 (2002); \textbf{89,} 011302 (2002); Phys. Rev. C\textbf{%
72}, 055502 (2005).

[3]\qquad KAMLAND Collaboration, T. Araki et al., Phys. Rev. Lett.\textbf{94}%
, 081801 (2005).

[4]\qquad Homestake Collaboration, T. Leveland et al. Astrophys. J. \textbf{%
496,} 505 (1998); GNO Collaboration, M. Altman et al. Phys. Lett. B\textbf{%
616}, 174 (2005); SAGE Collaboration, J. N. Abdurashitov et al., \ \ \ \ \ \
\ \ \ \ \ Nucl. Phys. Proc. Suppl. \textbf{110}, 315 (2002);
Super-KamiokandeCollaboration, J. Hosaka et al., Phys. Rev. D\textbf{73},
112001 (2006).

[5] S. M. Bilenky, C. Giunti and W. Grimus, Progress In Particle and Nuclear
Physics, \textbf{43}, 1 (1999).

[6] C. Giunti and M. Laveder, \textquotedblleft Neutrino
Mixing\textquotedblright ; hep-ph/0310238v2.

[7] B. Kayser, \textquotedblleft Neutrino Oscillation
Phenomenology\textquotedblright ; Proc. of the 61 st Scottish Universities
Summer School in Physics, Eds. C. Frogatt and P. Soler (to appear); arXiv:
0804.1121v3 [hep-ph].

[ [8] M. Fukugita and T. Yanagida, \textquotedblleft Physics of Neutrinos
and Applications to Astrophysics\textquotedblright\ (Springer Verlag Berlin
Heidelberg 2003).

[9] C. Giunti and C. W. Kim, \textquotedblleft Fundamentals of Neutrino
Physics and Astrophysics\textquotedblright\ (Oxford University Press, Oxford
2007).

[10] C. Giunti, \textquotedblleft Neutrino Flavor States and the Quantum
Theory of Neutrino Oscillations\textquotedblright\ (XI Mexican Workshop on
Particles and Fields, 7-12 November 2007)), arXiv: 0801. 0653 v1 [hep-ph].

11] C. Giunti, \textquotedblleft Fock States of Flavor Neutrinos are
Unphysical\textquotedblright ; Eur. Phys. J. C\textbf{39,} 377-382 (2005);
hep- ph/0312256v2.

\bigskip \lbrack 12] B. Pontecorvo, JETP \textbf{7, }172 (1958). \ \ 

[13] R. E. Schrock, Phys. Lett. \textbf{96B}, 159 (1980).

\ [14] Y. F. Li and Q. Y. Liu, JHEP 0610, 048 (2006).

[15] M. Blasone, A. Capolupo, F. Terranova and G. Vittiello, Phys. Rev. 
\textbf{D73}, 013003 (2005).

\ 16] S. M. Bilenky, F. von Feilitzsch and W. Potzel, J. Phys. G; Nucl.
Part. Phys. \textbf{34}, 987 (2007), hep- ph/0611285v2.

[17] M. Dvornikov, Phys. Lett. \textbf{B610}, 262 (2005), hep-ph/0411101.

[18] H. Fritzsch, "Flavor, Mixing, Neutrino Masses and Neutrino
Oscillations", arXiv: 0902.2817.

[19] P. F. Harrison, D. H. Perkins and W. G. Scott, Phys. Lett. \textbf{%
B530, }167 (2002).

\bigskip\ \ \ \ \ \ \ \ \ \ \ \ \ \ \ \ \ \ \ \ \ \ \ \ \ \ \ \ \ \ \ \ \ \
\ \ \ \ \ \ \ \ \ \ \ \ \ \ \ \ \ \ \ \ \ \ \ \ \ \ \ \ \ \ \ \ \ \ \ \ \ \
\ \ \ \ \ \ \ \ \ \ \ \ \ \ \ \ \ \ \ \ \ \ \ \ \ \ \ \ \ \ \ \ \ \ \ \ \ \
\ \ \ \ \ \ \ \ \ \ \ \ \ \ \ \ \ \ \ \ \ \ \ \ \ \ \ \ \ \ \ \ \ \ \ \ \ \
\ \ \ \ \ \ \ \ \ \ \ \ \ \ \ \ \ \ \ \ \ \ \ \ \ \ \ \ \ \ \ \ \ \ \ \ \ \
\ \ \ \ \ \ \ \ \ \ \ \ \ \ \ \ \ \ \ \ \ \ \ \ \ \ \ \ \ \ \ \ \ \ \ \ \ \
\ \ \ \ \ \ \ \ \ \ \ \ \ \ \ \ \ \ \ \ \ \ \ \ \ \ \ \ \ \ \ \ \ \ \ \ \ \
\ \ \ \ \ \ \ \ \ \ 

.

\bigskip

\ \ \ \ \ \ \ \ \ \ \ \ \ \ \ \ \ \ \ \ \ \ \ \ \ \ \ \ \ \ \ \ \ \ \ \ \ \
\ \ \ \ \ \ \ \ \ \ \ \ \ \ \ \ \ \ \ \ \ \ \ \ \ \ \ \ \ \ \ \ \ \ \ \ \ \
\ \ \ \ \ \ \ \ \ \ \ \ \ \ \ \ \ \ \ \ \ \ \ \ \ \ \ \ \ \ \ \ \ \ \ \ \ \
\ \ \ \ \ \ \ \ \ \ \ \ \ \ \ \ \ \ \ \ \ \ \ \ \ \ \ \ \ \ \ \ \ \ \ \ \ \
\ \ \ \ \ \ \ \ \ \ \ \ \ \ \ \ \ \ \ \ \ \ \ \ \ \ \ \ \ \ \ \ \ \ \ \ \ \
\ \ \ \ \ \ \ \ \ \ \ \ \ \ \ \ \ \ \ \ \ \ \ \ \ \ \ \ \ \ \ \ \ \ \ \ \ \
\ \ \ \ \ \ \ \ \ \ \ \ \ \ \ \ \ \ \ \ \ \ \ \ \ \ \ \ \ \ \ \ \ \ \ \ \ \
\ \ \ \ \ \ \ \ \ \ \ \ \ \ \ \ \ \ \ \ \ \ \ \ \ \ \ \ \ \ \ \ \ \ \ \ \ \
\ \ \ \ \ \ \ \ \ \ \ \ \ \ \ \ \ \ \ \ \ \ \ \ \ \ \ \ \ \ \ \ \ \ \ \ \ \
\ \ \ \ \ \ \ \ \ \ \ \ \ \ \ \ \ \ \ \ \ \ \ \ \ \ \ \ \ \ \ \ \ \ \ \ \ \
\ \ \ \ \ \ \ \ \ \ \ \ \ \ \ \ \ \ \ \ \ \ \ \ \ \ \ \ \ \ \ \ \ \ \ \ \ \
\ \ \ \ \ \ \ \ \ \ \ \ \ \ \ \ \ \ \ \ \ \ \ \ \ \ \ \ \ \ \ \ \ \ \ \ \ \
\ \ \ \ \ \ \ \ \ \ \ \ \ \ \ \ \ \ \ \ \ \ \ \ \ \ \ \ \ \ \ \ \ \ \ \ \ \
\ \ \ \ \ \ \ \ \ \ \ \ \ \ \ \ \ \ \ \ \ \ \ \ \ \ \ \ \ \ \ \ \ \ \ \ \ \
\ \ \ \ \ \ \ \ \ \ \ \ \ \ \ \ \ \ \ \ \ \ \ \ \ \ \ \ \ \ \ \ \ \ \ \ \ \
\ \ \ \ \ \ \ \ \ \ \ \ \ \ \ \ \ \ \ \ \ \ \ \ \ \ \ \ \ \ \ \ \ \ \ \ \ \
\ \ \ \ \ \ \ \ \ \ \ \ \ \ \ \ \ \ \ \ \ \ \ \ \ \ \ \ \ \ \ \ \ \ \ \ \ \
\ \ \ \ \ \ \ \ \ \ \ \ \ \ \ \ \ \ \ \ \ \ \ \ \ \ \ \ \ \ \ \ \ \ \ \ \ \
\ \ \ \ \ \ \ \ \ \ \ \ \ \ \ \ \ \ \ \ \ \ \ \ \ \ \ \ \ \ \ \ \ \ \ \ \ \
\ \ \qquad

\end{document}